# REFLECTIONLESS TUNNELLING OF LIGHT IN GRADIENT OPTICS

**Alexander B. SHVARTSBURG [a] and Guillaume PETITE [b]**

(a) Central Design Bureau for Unique Instrumentation of the R A S, Butlerov Str. 15, 117342, Moscow, Russia

(b) Laboratoire des Solides Irradiés, CEA/DSM/DRECAM , CNRS (UMR 7642) et Ecole Polytechnique, 91128, Palaiseau, France

Abstract : We analyse the optical (or microwave) tunnelling properties of electromagnetic waves passing through thin films presenting a specific index profile providing a cut-off frequency, when they are used below this frequency. We show that contrary to the usual case of a square index profile, where tunnelling is accompanied with a strong attenuation of the wave due to reflection, such films present the possibility of a reflectionless tunnelling, where the incoming intensity is totally transmitted.

Tunnelling of EM waves through the opaque barriers in optical and microwave ranges had been studied intensively during the last decades.[1-5] Analysis of these phenomena, including percolation of the attenuated energy flow of evanescent waves through the non-transparent barrier and phase saturation of these waves (Hartman effect[6,7]), was usually based on the model of rectangular potential barrier, pioneered by Gamov for the theory of $\alpha$- decay as long ago as in 1928.[8] This approach was used for the treatment of evanescent modes in bi-prism systems,[9-12] and undersized waveguides.[7,13,14] However, the observation of these phenomena is usually impeded due to strong attenuation of tunnelling waves, while travelling through barriers, accompanied by a strong reflection of waves from such barriers. This paper is devoted to the possibility of reflectionless tunnelling of EM wave through specially shaped barriers with waveguide-like dispersion,[15] when the wave's frequency is less than their cut-off frequency.

To demonstrate the possibility of reflectionless tunnelling, let us consider a linearly-polarized EM wave with components $E_x$ and $H_y$, propagating in $z$-direction, incidenting perpendicularly to the interface of an optical system containing $m$ ($m \geq 1$) thin heterogeneous dielectric films, located, for mechanical support, on an homogeneous transparent plate (substrate). Supposing all the layers to be lossless and non-magnetic and expressing the field components through the vector-potential $A$ ($A_x = \psi$, $A_y = A_z = 0$) so, that

$$E_x = -\frac{1}{c}\frac{\partial \psi}{\partial t}, \qquad H_y = \frac{\partial \psi}{\partial z} \qquad (1)$$

one can reduce the system of Maxwell equations, related to this geometry, to one equation, governing the function $\psi$ in each layer:

$$\frac{\partial^2 \psi}{\partial z^2} - \frac{\varepsilon(z)}{c^2}\frac{\partial^2 \psi}{\partial t^2} = 0 \qquad (2)$$

Here $\varepsilon(z)$ is the dielectric susceptibility : for the homogeneous plate $\varepsilon = n^2$, where $n$ is the refractive index. For the heterogeneous films $\varepsilon$ is characterized by coordinate-dependent profile $\varepsilon(z)=n_0^2 U^2(z)$; where $n_0$ is the value of refractive index on the interfaces of heterogeneous films, and the function $U(z)$ presents the symmetric concave profile

$$U(z) = \left(1 + \frac{z}{L_1} + \frac{z^2}{L_2}\right)^{-1} \tag{3}$$

Here $L_1$ and $L_2$ are the spatial scales, connected with the film's thickness $d$ and minimal value of $\varepsilon$: $\varepsilon_m = n_0^2 U_m$:

$$L_1 = \frac{d}{4y^2}; \quad L_2 = \frac{d}{2y}; \quad y = \sqrt{U_m^{-1} - 1} \tag{4}$$

The values of $d$ and $U_m$ in this model remain arbitrary. To find the reflection coefficient let us consider first the reflectivity of one heterogeneous film with profile (2), supported by the substrate. This profile was shown to possess a non-local heterogeneity - induced dispersion and a cut - off frequency $\Omega$,[15]

$$\Omega = \frac{2cy\sqrt{1+y^2}}{n_0 d} \tag{5}$$

Eq. (2) has different solutions in the spectral ranges $\omega > \Omega$ and $\omega < \Omega$, related to the travelling and tunnelling cases respectively; considering here the case of a tunnelling wave and using the corresponding solution of (2) that we derived in a previous paper,[16] we will present the function $\psi = \psi_1$ as a sum of forward and backward waves :

$$\psi_1 = \frac{A_1}{\sqrt{U(z)}}[\exp(-p\eta) + Q_1 \exp(p\eta)] \tag{6}$$

$$p = \frac{\omega N_-}{c}; \quad N_-^2 = u^2 - 1; \quad u = \frac{\Omega}{\omega}$$

$$\tag{7}$$

$$\eta = \frac{L_2}{\sqrt{1+y^2}} \ln\left[\frac{1 + y_+ z/2L_2}{1 - y_- z/2L_2}\right]; \quad y\pm = \sqrt{1+y^2} \pm y$$

The coexistence in the film of two different waves, an evanescent wave whose intensity decreases with increasing $z$, and an "antievanescent" one whose intensity on the contrary increases for increasing $z$ (which would respectively correspond to a forward and a backward propagating wave if the film was transmitting) is an essential feature which is at the heart of the physics described hereafter.

In the homogeneous plate with a refractive index $n$, function $\psi = \psi_2$ may be written as

$$\psi_2 = A_2[\exp(ikz) + Q_2 \exp(-ikz)]; \quad k = \frac{\omega n}{c} \tag{8}$$

Quantities $Q_1$ and $Q_2$ determine the amplitudes of backward waves. The constants $A_1$, $A_2$ and $Q_1$, $Q_2$ have to be found from the continuity conditions for the field components $E_x$ and $H_y$. Substitution of functions $\psi_1$ and $\psi_2$ into (1) brings the values $E_x$ and $H_y$. Let us first consider the wave incidenting on the first interface of the heterogeneous film (plane $z = 0$). The continuity conditions yield the reflection coefficient $R$

$$R = \left[1 + \frac{i\gamma}{2} - in_0 N_- \frac{1-Q_1}{1+Q_1}\right]\left[1 - \frac{i\gamma}{2} + in_0 N_- \frac{1-Q_1}{1+Q_1}\right]^{-1} ; \qquad (9)$$

To find the quantity $Q_1$, linked with $Q_2$, one can use the continuity conditions on the interface between the film and the substrate, resulting in :

$$Q_1 = \frac{S_1}{S_2} \exp(-2p\eta_0)$$

$$S_1 = n_0 N_- + \frac{\gamma}{2} + iM ; \qquad S_2 = n_0 N_- - \frac{\gamma}{2} - iM ; \qquad M = n\frac{1-Q_2}{1+Q_2} \qquad (10)$$

$$\gamma = \frac{2yun_0}{\sqrt{1+y^2}} ; \qquad \eta_0 = \eta(d) = \frac{L_2}{\sqrt{1+y^2}} \ln\left(\frac{y_+}{y_-}\right)$$

The quantity $Q_2$, characterizing the wave field inside the substrate (thickness D ) can be found from the continuity conditions on the air-substrate interface:

$$Q_2 = \frac{n-1}{n+1} \exp(2i\omega n D/c) \qquad (11)$$

Using (11), we find $M$ (10):

$$M = n\frac{1-int_1}{n-it_1} \quad ; \quad t_1 = tg(\omega nD/c) \tag{12}$$

Substitution of (11) and (12) into (10) and (9) brings the value of $Q_1$ and, finally, the reflection coefficient $R$ for arbitrary parameters of substrate $n$ and $D$. It is remarkable, that in the special case where the substrate's thickness contains an integer amount of half-waves

$$kD = s; \quad s = 1; 2; 3\ldots, \tag{13}$$

$M$ is equal to unity; as in the case of a free-standing film ($n = 1$). In both cases the values $S_1$ and $S_2$ (10), as well as the values of reflection coefficient $R$ (9) coincide :

$$R = \frac{th(p\eta_0)\left[1+\frac{\gamma^2}{4}+n_0^2 N_-^2\right] - \gamma n_0 N_-}{th(p\eta_0)\left[1-\frac{\gamma^2}{4}-n_0^2 N_-^2\right] + \gamma n_0 N_- + i(2n_0 N_- - \gamma th(p\eta_0))} \tag{14}$$

Thus, in case (13) the substrate has no influence on the tunnelling process; namely this case will be discussed below. When the heterogeneity is vanishing ($\gamma \to 0, \eta_0 \to d$) eq. (13) is reduced to a well known formula for the reflection coefficient of a plasma-like layer for frequencies below the plasma frequency

$$R = \frac{th(pd)(u^2)}{th(pd)(2-u^2)+2iN_-} \tag{15}$$

It is essential that the tunnelling through the rectangular barrier (15) is always accompanied by some reflection (coefficient $R$ (15) is never equal to zero), meanwhile the value $R$ for concave barrier (14) may reach the zero value. Before examining this reflectionless tunnelling it is worthwhile to generalize eq. (14) for a stack, containing *m* similar films, and located on the substrate, obeying the condition (13). Considering again the continuity conditions on each interface between the neighbouring films, we find that this generalization implies in replacement of factor $th(p\eta_0)$ in (14) on the factor $(mp\eta_0)$. Making this replacement, one can write the condition of reflectionless tunnelling in a form:

$$th(mp\eta_0) = \frac{\gamma n_0 N_-}{1 + \frac{\gamma^2}{4} + n_0^2 N_-^2} \tag{16}$$

Considering the transmission function $T = |T|\exp(i\phi_t)$, we find from the relation $|R|^2 + |T|^2 = 1$ the value $|T| = 1$, which indicates total transmission. The cancellation of the reflected wave results from the superposition of the wave reflected at the interface $z=0$ and the backward transmitted part of the anti-evanescent wave. It is subject to both a phase and an amplitude condition on the anti-evanescent wave. In the case of a square tunneling barrier, the phase and the amplitude of this wave are linked in such a way that this cancellation cannot be obtained. Here, the heterogeneity induced dispersion of the film can be viewed as an additional degree of freedom which allows to realize this compensation.

The phase shift of the tunnelling wave $\phi_t$ can be obtained from (6), in which the amplitude $A_2$ has to be calculated by means of continuity conditions. Omitting some tedious algebra, one can present the phase shift $\phi_t$ in a form

$$tg(\phi_t) = \frac{\gamma}{1 + n_0^2 N_-^2 - \frac{\gamma^2}{4}} \tag{17}$$

To find the parameters of optical system characterized by eq.(16), for some frequency $\omega$, the depth of modulation (parameter $y$ (4)) and the value $n_0$ being known, it is worthwhile to introduce the variable $x = \sqrt{1-u^{-2}}$ and to rewrite eq. (16) in a form

$$th(mlx) = \frac{2xy}{\sqrt{1+y^2}} \left[ \frac{1}{n_0^2} + \frac{y^2}{1+y^2} + x^2\left(1 - \frac{1}{n_0^2}\right) \right]^{-1} ; \quad l = \ln\left(\frac{y_+}{y_-}\right); \quad (18)$$

Solving this equation with respect to $x$, we will find the variable $u$ and the cut-off frequency $\Omega = \omega u$. Finally, making use of (5), one can calculate the film thickness $d$.

Let us give a numerical example. Considering, e.g., the wavelength $\lambda = 800$ nm, tunnelling without reflection through a stack containing two films characterized by a thickness $d = 65$ nm, $n_0 = 2.35$, depth of modulation of refractive index 25% (y = 0.577, $n_{min}$ = 1.76), we will find the values $\Omega = 2.6 \; 10^{15}$ rad sec$^{-1}$, $u$ = 1.1; the phase shift is $\phi_t = 1.38$ rad. Thus, the zero-crossing point is traversing this distance $2d$ during the subluminal time $t = \phi_t / \omega = 0.585$ fs = 1.36 $t_0$, where $t_0$ = $2d/c$ relates to propagation in air; however, the phase shift, obtained for this free propagation, would be 1.02 rad, which is less than the abovementioned tunnelling phase shift 1.34 rad. One obtains here the result already mentioned (or observed) by several authors[17-20] that despite the fact that the "phase time" seems to indicate a superluminal propagation, the signal velocity is always subluminal, confirming a result previously obtained on the grounds of an independent calculation of the group velocity of the evanescent wave in such inhomogeneous films.[16]

As a conclusion, we have shown that some gradient index films presenting a cut-off frequency – provided by their heterogeneity-induced dispersion – and used in the tunnelling regime (below their cut-off frequency) could yield reflectionless tunnelling, i.e. tunnelling without attenuation of the transmitted wave, a property that is never exhibited by regular constant-index films. The above calculation can be extended to semi-infinite substrates (at the cost of an increased complexity) and, more important, to inclined incidence. Moreover, since the above theory is scalable simply by a change of the geometrical scales $L_1$ and $L_2$, it can be applied to any wavelength scales, from the tunnelling of electronic wavefunctions in heterostructures to that of microwaves in waveguides, provided the material can be realized. In the optical domain, gradient index films are currently synthesized (e.g. oxy-nitride films with composition gradients) possessing all the required characteristics (well controled gradient of significant amplitude, together with a low intrinsic material dispersion) for operation in the near IR range. Likewise, gradient index materials are used in some RF applications, making RF experiments of reflectionless tunneling a realistic objective. On the contrary, obtaining an equivalent control at the nanometer scale, which would be necessary for some quantum physical applications (like tunneling of particles), may be much more difficult, so that electromagnetism appears the best test case for the ideas exposed hereabove.

The authors acknowledge the help of NATO (grant n° PST.CLG 980331).

References.

1. E. H. Hauge and J. A. Stovneng, Rev. Mod. Phys. **61**, 917 – 935 ( 1989 ).

2. R. Y. Chiao and A. M .Steinberg, "Tunneling times and superluminality", in "*Progress in Optics*", ed. by E. Wolf, Vol. 37, 345 - 405 (1997).

3. V. S. Olkhovsky, E. Recami and J. Jakiel, "Unified time analysis of photon and particle tunneling", Physics Reports, Vol. 398, 133 – 178 (2004).

4. F. Fornel, "Evanescent waves: from Newtonian optics to atomic optics", Springer Series in Optical Sciences, Vol. 73, (2001).

5. A. Enders and G. Nimtz, J. Phys. (France) I, **2**, 1963 (1992), Phys. Rev. E, **48**, 652 (1993)

6. T. E. Hartman, J. Appl. Phys., **33**, 3427 – 3433 (1962).

7. V. S. Olkhovsky and E. Recami, Phys. Rep., **214**, 339-356 (1992)

8. G. A. Gamow, Z. Phys., **51**, 204 – 210 (1928).

9. B. Lee and W. Lee, J. Opt. Soc. Am. B **14**, 777 – 781 (1997).

10. Chun-Fang Li and Qi Wang, J. Opt. Soc. Am. B **18**, 1174 – 1179 (2001).

11. V. Laude and P. Tournois, J. Opt. Soc. Am. B **16**, 194 – 198 (1999).

12 A. Haibel, G. Nimtz and A. A. Stahlhofen, Phys. Revs. E., 63, 047601 (2001)

13 A. Pablo L. Barbero, H.E. Fernandez-Figueroa and E . Recami, Phys. Rev. E., 62, 8628 (2000)

14. A. Ranfagni, P. Fabeni, G. Pazzi and D. Mugnai, Phys. Rev. E, **48**, 1453 – 1458 (1993).

15. A. Shvartsburg, G. Petite and P. Hecquet, J. Opt. Soc. Am. A **17**, 2267 – 2272 (2000).

16. A. Shvartsburg and G. Petite, European Phys. J. D, **36**, 111-118 (2005).

17. V.S. Olkhovsky, E. Recami and G. Salesi, Europhys. Lett. **57,** 879-884,(2002);

18 V.S. Olkhovski, E. Recami and A.K. Zaichenko, Europhys. Lett. **70,** 712-718 (2005)

19 Y. Aharonov, N. Erez and B. Reznik, Phys. Rev. A, **65,** 052124 (2002)

20 S. Longhi, P. Laporta, M. Belmonte and E. Recami, Phys. Rev. E, **65,** 046610 (2002)